# TaskMe: Multi-Task Allocation in Mobile Crowd Sensing

**Yan Liu[†], Bin Guo[†], Yang Wang[††], Wenle Wu[†], Zhiwen Yu[†], Daqing Zhang[†††]**
[†] Northwestern Polytechnical University, Xi'an 710129, Shaanxi, China
[††] Shenzhen Institutes of Advanced Technology, Shenzhen 518055, China
[†††] TELECOM SudParis, Evry 91011, France
guob@nwpu.edu.cn

### ABSTRACT
Task allocation or participant selection is a key issue in Mobile Crowd Sensing (MCS). While previous participant selection approaches mainly focus on selecting a proper subset of users for a single MCS task, multi-task-oriented participant selection is essential and useful for the efficiency of large-scale MCS platforms. This paper proposes TaskMe, a participant selection framework for multi-task MCS environments. In particular, two typical multi-task allocation situations with bi-objective optimization goals are studied: (1) For FPMT (few participants, more tasks), each participant is required to complete multiple tasks and the optimization goal is to maximize the total number of accomplished tasks while minimizing the total movement distance. (2) For MPFT (more participants, few tasks), each participant is selected to perform one task based on pre-registered working areas in view of privacy, and the optimization objective is to minimize total incentive payments while minimizing the total traveling distance. Two optimal algorithms based on the Minimum Cost Maximum Flow theory are proposed for FPMT, and two algorithms based on the multi-objective optimization theory are proposed for MPFT. Experiments verify that the proposed algorithms outperform baselines based on a large-scale real-word dataset under different experiment settings (the number of tasks, various task distributions, etc.).

### Author Keywords
Mobile crowd sensing; multi-task allocation; participant selection; bi-objective optimization.

### INTRODUCTION
Mobile crowd sensing (MCS) [11,13] is a new way of sensing in which a crowd of mobile users utilize their smart devices to conduct complex computation and large-scale sensing tasks. It has stimulated a lot of attractive applications, such as air quality monitoring [33], traffic information mapping [9], public information sharing [12], and so on. There are a series of things needed to be done in MCS, such as publishing sensing tasks, selecting users to complete tasks, collecting data, etc. Therefore, the development of a generic MCS platform is indispensable for a variety of MCS tasks. There have been some MCS platforms developed, like Campaignr [4] and Medusa [22]. However, most platforms are used mainly for task publishing and data collection. Moreover, users in these platforms decide which tasks to complete by themselves, instead of being selected to match with suitable tasks.

Participant selection problem is one of the major challenges in the MCS platform, which has an impact on the efficiency and quality of the tasks. There have been recently numerous studies on MCS participant selection [6,15,27,31]. In these works, they select an optimal set of users to complete a task while fulfilling some needs or constraints of the task (e.g., the budget cost). However, existing works are mostly single-task oriented, and they do not consider the task allocation problem in a large-scale MCS platform (e.g., at the city-scale) where there can be multiple concurrent tasks published by different requesters. Therefore, we are faced with new challenges in platform-oriented multi-task allocation. This paper aims to address these new challenges by studying multi-task allocation in MCS platforms.

For participant selection, the relationship between the number of active participants and the number of tasks to be completed in the MCS platform is a significant factor that influences task completion. This, however, has rarely been studied in existing works. Moreover, most studies select participants based on the single optimization objective (e.g., maximizing the number of accomplished tasks), but it is necessary to take various optimization objectives (e.g., maximizing the number of accomplished tasks and minimizing total incentive payments) into consideration in many cases. In addition, privacy is a serious problem in participant selection. However, most works obtain the precise location of mobile users to complete location-based tasks without considering privacy. Considering of these issues, two scenarios of multi-task allocation are presented below to illustrate the practical values of studying this problem, as shown in Fig. 1.

There is a set of volunteers registered in the MCS platform. The platform can collect information (e.g., location) from them for task allocation. Task publishers offer a variety of sensing tasks, like traffic dynamics sensing, environment monitoring, etc. Considering that most sensing tasks are likely to arrive one after the other, the MCS platform could allocate multiple sensing tasks during the same time span, which are published in the platform within a period of time (e.g., 15 minutes). There will be two situations regarding the number of tasks and active participants, which leads to different multi-task allocation problems. The first one is that there are more tasks to be completed, while only few participants are available, and we call it the "FPMT problem". The second one is just the opposite, where the participant resources are rich while tasks are not that many, named the "MPFT problem".

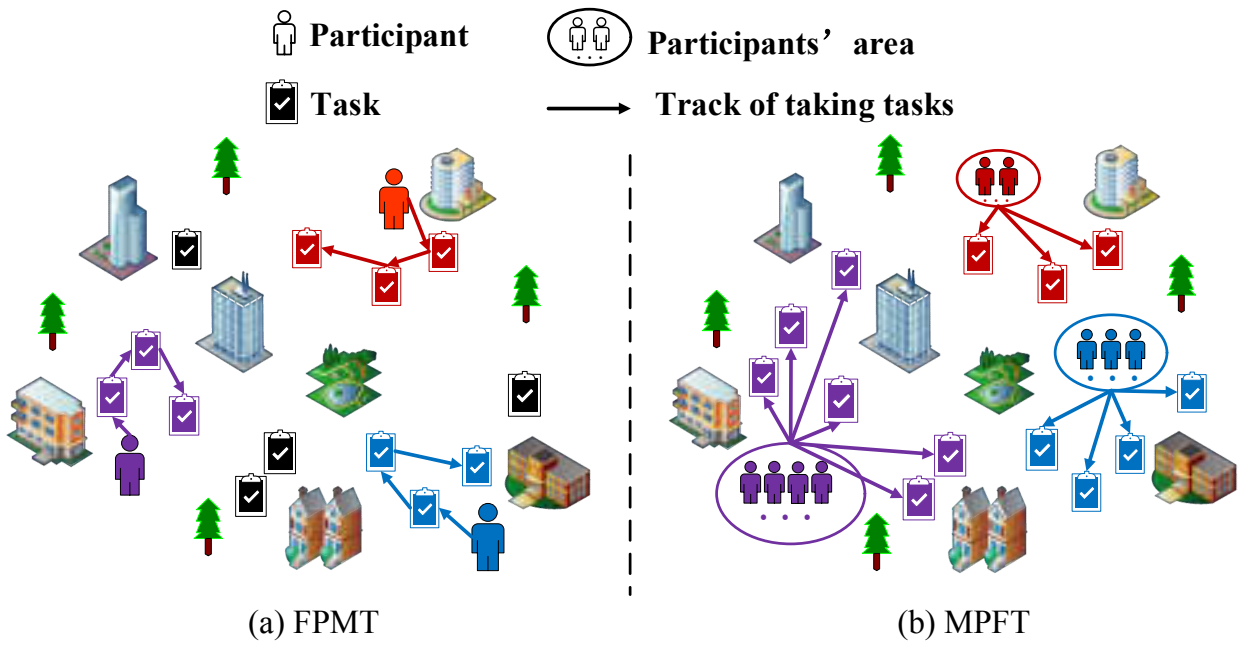

**Figure 1. Two scenarios of participant selection.**

***FPMT problem****. For example, there are few people on the road during a heavy rain. Though, many urgent tasks are published in the MCS platform, such as collecting traffic dynamics information, monitoring drainage status, etc. In order to complete tasks as many as possible, each participant is required to perform multiple tasks. In addition, participants need to move from current location to the location of tasks to complete them. The traveling distance thus becomes the primary factor to reduce the completion time of such urgent tasks. On balance, the objectives of the platform are not only to maximize the number of completed tasks, but also to minimize the total traveling distance. The incentive and privacy are not the main factors in FPMT problem owing to the emergency of the tasks and poor participant resources.*

***MPFT problem****. Unlike the emergency in FPMT problem, some tasks do not need to be performed immediately, such as measuring air quality, collecting information of public facility, etc. Therefore, the platform can allocate these tasks when there are rich participants, and each participant only needs to perform one task to improve the quality of task performing. Moreover, as for non-urgent tasks, to protect user privacy, the platform allows participants to register region-level interested areas (e.g., a 2km*2km area or a university campus) for task performing, rather than requiring their precise location. We assume that the incentive of each participant is in inverse proportion with the number of users in this area (i.e., the market principles). The platform then aims to minimize total incentive payments for participants and minimize traveling distance to complete tasks.*

This paper presents TaskMe, a framework for multi-task allocation. The main contributions of it are presented below:

(1) We investigate the impact of different factors to multi-task allocation and formulate two bi-objective optimization problems under different situations: the FPMT problem aims to maximize the number of accomplished tasks and minimize the total traveling distance, and the MPFT problem aims to minimize total incentive payments and minimize the total movement distance.

(2) We transform the FPMT problem using the Minimum Cost Maximum Flow (MCMF) theory [14], and construct a new MCMF model for the FPMT problem by considering different constraints, then propose the MT-MCMF and MTP-MCMF algorithms for FPMT. The MPFT problem is addressed by the multi-objective optimization theory [25] with transforming bi-objective optimizations to single-objective optimizations. Two algorithms, W-ILP (based on the linear weight method) and C-ILP (based on the constraint method) are proposed. Compared to existing algorithms, our algorithms are characterized by their distinct features in optimizing multiple objectives at the same time while still paying less cost.

(3) Evaluating the performance of the proposed algorithms of FPMT and MPFT, and conducting experiments based on real-world datasets D4D [2], which contains call records of 50,000 mobile users. The results indicate that the proposed algorithms perform better than the baseline approaches based on greedy heuristics.

## RELATED WORK

There have been numerous human-powered platforms developed. Mturk [3] is a well-known crowdsourcing platform, which has attracted a lot of requesters to publish tasks and plenty of participants to complete tasks. However, Mturk tasks are generally performed online while not in the real-world. As opposed to online crowdsourcing, MCS platform allows task requesters to publish location-based tasks. Representative examples include AndWellness [17], PRISM [10], Campaignr [4], Medusa [22], etc. However, participants in these platforms decide which tasks to complete by themselves, instead of being selected by the platform based on different sensing purposes.

Participant selection is a key research issue in MCS and has attracted much attention in recent years. Reddy *et al.* [23] studied a recruitment framework to identify appropriate participants for data collections based on geographic and temporal availability. Zhang *et al.* [32] proposed a novel participant selection framework, aiming to minimize incentive payments by selecting a small number of users with satisfying probabilistic coverage constraint. Cardone *et al.* [5] studied how to select participants to maximize the spatial coverage of crowdsensing. However, participants in these works only are required to complete one task, which cannot fully utilize the resources of mobile users, especially in the case of insufficient resources. Privacy issues are also not considered in these studies.

Quite few efforts are made to multi-task allocation. Xiao *et al*. [29] studied the task assignment problem in mobile social networks, and their objective is to minimize the average makespan. Song *et al*. [24] proposed a multitask-oriented participant selection strategy to select a minimum subset of participants to satisfy the quality-of-information requirements of concurrent tasks being serviced with total budget constraints. However, the relationship between the number of tasks and participants is not studied. In addition, there are few studies about multi-objective optimization in MCS task allocation. Xiong *et al*. [30] selected participants with the aim to maximize the coverage quality of the sensing task while satisfying the incentive budget constraint, but their work is not about multi-task allocation.

In our paper, we focus on multi-task allocation in the MCS platform under two typical situations, and study two bi-objective optimization problems for participant selection.

## TASK ALLOCATION OF FPMT PROBLEM

### Problem Analysis

Due to poor participant resources and the emergency of tasks, the objectives of FPMT are not only to maximize the total number of accomplished tasks, but also to minimize the total movement distance of participants. There are a number of location-based sensing tasks and some active participants in the MCS platform. For the sensing tasks, several participants are needed to perform each task to enhance the quality of sensing data. We also set an upper value for it to avoid obtaining much redundant data. For the participants, the platform requires each of them to perform multiple tasks to maximize the number of accomplished tasks. Considering of fairness and completion quality, we set the same values for the number of tasks allocated to each participant. For the platform, because most tasks in FPMT are urgent, the platform thus needs to obtain precise locations of participants to improve the efficiency of tasks completion without regarding user privacy. Obviously, the total number of accomplished tasks is the main optimization goal for task publishers to improve task accomplishment ratio. What's more, participants need to complete tasks as soon as possible due to the emergency request. We simply assume that the travelling time is proportional to the movement distance, minimizing the total movement distance thus becomes the other optimization goal of the FPMT problem.

Given a crowd of active participants in the MCS platform, denoted by the set $U=\{u_1,u_2...u_i...u_m\}$. Supposing that each of them is required to complete $q$ tasks. In addition, we assume that there are a variety of sensing tasks published by task requesters, denoted by $T=\{t_1,t_2...t_j...t_n\}$. Each task $t_j$ will be assigned to at most $p_j$ users, and we use $UT_j=\{u_{j1},u_{j2},u_{j3}...\}$ to denote the participant set who finish task $t_j$. Moreover, we use $TU_i=\{t_{i1},t_{i2},t_{i3}...\}$ to denote the set of tasks assigned to participant $u_i$, and $D(TU_i)$ is the total movement distance of $u_i$ to complete tasks in $TU_i$. Then, the FPMT problem can be formulated as Eq. (1)-(4):

$$\text{maximize} \sum_{i=1}^{m} |TU_i| \quad (1)$$

$$\text{minimize} \sum_{i=1}^{m} D(TU_i) \quad (2)$$

Subject to:

$$|TU_i| = q \ (1 \le i \le m) \quad (3)$$

$$|UT_j| \le p_j \ (1 \le j \le n) \quad (4)$$

There are two main challenges to solve FPMT: one is that each participant needs to complete multiple tasks; the other is that there are two optimization objectives in FPMT. For the first challenge, it is the dynamic process for multi-task completion compared to single-task owing to location-based tasks. FPMT includes not only the combinational optimization problem [8] between the tasks and participants, but also the shortest path problem [16] for each participant. It is difficult to find an algorithm to obtain the optimal solution of FPMT simultaneously taking into account both the combinational optimization and the shortest path. On the contrary, it is perhaps a feasible way that we could figure out all possible task combinations for each participant in advance, and obtain optimal solution of the combinational optimization problem. For the second challenge, two optimization objectives are contradictory, and it is impossible to obtain optimal solution to meet requirements of two optimization objectives simultaneously. There, however, could be a tradeoff between the number of accomplished tasks and the total movement distance.

According to the analysis of FPMT, we adopt the MCMF model to solve it. MCMF model is the classic method to solve bi-objective optimization problem. We transform FPMT to the MCMF problem, and construct a new MCMF model by considering different constraints.

### MT-MCMF Algorithm

The MCMF model aims to find a group of optimal paths with minimum cost and maximum flow. Therefore, we transform FPMT to the MCMF problem, where the total movement distance represents the cost and the total number of accomplished tasks is modeled as the flow. Note that the flow network of the MCMF model is a directed graph, where each edge has capacity, flow and cost [28]. The capacity of an edge represents the maximum amount of flow that can pass through an edge. However, the MCMF model cannot be used directly and there are several problems to be solved when using it: (1) it is impossible to have only participant nodes and task nodes in the flow network; (2) it is a difficult problem to show the paths of each participant with $q$ tasks; (3) we need to represent different requirements of participants and tasks in the flow network. Considering the three issues mentioned above, we have reconstructed the MCMF model, and made three improvements to solve them, as shown in Fig. 2.

*(1) Adding the source node and sink node to the flow network.* For the edges between the source and participant nodes, the capacity of each edge is $q$, which reflects that each participant is required to perform $q$ tasks; the cost of the edge is 0, because that the edges between the source and participant nodes are just used for inflowing flow with no movement distance. In addition, the flow inflowing to the sink from the participant and task nodes should reflect the completion of tasks. For the sink, the total maximum flow of sink is $n \times p$, where the capacity of each edge between the task node and the sink is $p$, and the cost of the edge is 0.

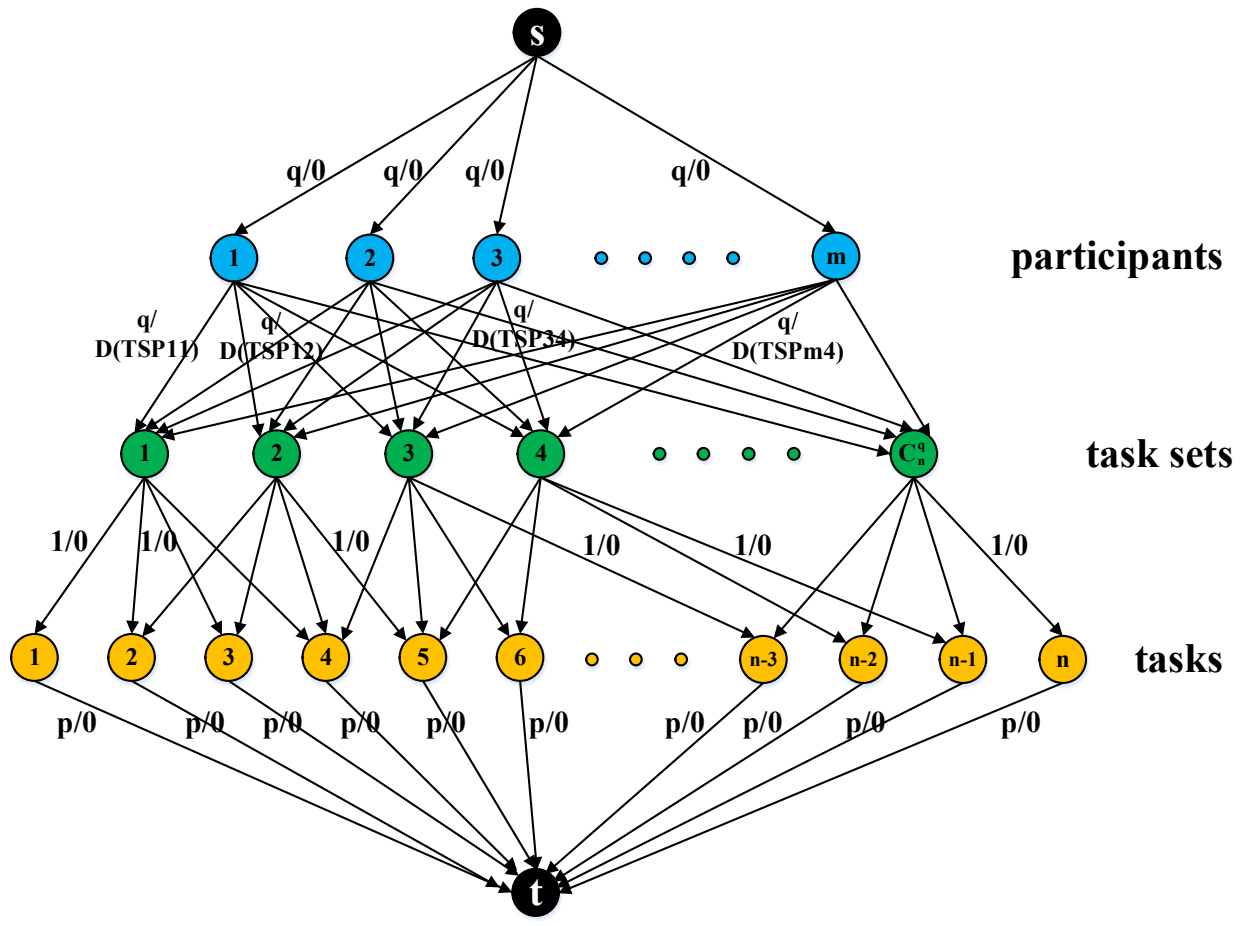

**Figure 2. MCMF model of FPMT problem.**

*(2) Adding the second-level nodes (green nodes) to the flow network to represent task sets.* We enumerate all possible task sets in advance and each task set has $q$ tasks picked from all $n$ tasks. Specifically, there are a total of $C_n^q$ task sets, and any task set could be performed by a participant. Moreover, the capacity of the edge between the first-level node and the second-level node is $q$. In addition, because task sets in the flow network cannot show the sequence of accomplished tasks, we need to compute the shortest path between the participant and the tasks in the task set in advance, like solving the TSP. Therefore, the cost of the edge between the participant node and the task set node is the distance of TSP including the participant and the tasks in the task set.

*(3) Adding the third-level nodes (yellow nodes) to show the completion of tasks.* We set different capacities of edges to satisfy different requirements of participants and tasks. We assume that one task set could be performed by only one participant. In addition, the capacity of the edge between the task node and the sink is $p$, because a task could be performed by at most $p$ participants.

In general, there are two main methods to solve the MCMF problem: the first one keeps the feasibility of the problem with maximum flow; the second one keeps the optimality of the problem with minimum cost. The FPMT problem is that the participant resources are deficient to the tasks, which means that the flows of the edges between the source and participant nodes are full of maximum flow in the flow network according to the proposition (proved as follows). Based on this, we adopt the second method to minimize the total movement distance. However, there is another important problem to be solved. Each task set node includes $q$ tasks, and the flow of the path between the participant node and task set node can be 0 or $q$ corresponded with the TSP, which cannot be adjusted incrementally according to the second method. Based on the above analysis, we propose the MT-MCMF algorithm for FPMT.

***Proposition.*** The flows of the edges between the source and participant nodes are full if the participant resources are deficient to the tasks.

***Proof.*** As we can see in the Fig. 2, the maximum flows of the edges between the source and participant nodes are $m \times q$. We assume that the edges between source and participant nodes of the final solution are filled with the flow less than the maximum flow, this implies that there is no augmenting path (a path from the source to sink in the residual network, which is the network pruning the edges filled with the flow) in flow network according to the theory [14]. Because that the flow of the edge between the source and the participant node is 0 or $q$, there must be at least an edge between the source and the participant nodes with no flow. Therefore, there must be at least an augmenting path in the flow network based on more tasks, which is incompatible with hypothesis.

**Algorithm 1**: MT-MCMF

**Input**: the user set $U$, the task set $T$

**Output**: the selected participant set $P$, the corresponding task set $T'$

1: select $C_n^q$ task-task sets $TT$ from task set $T$
2: form $|U| \times C_n^q$ participant-task sets $PT$ by adding each participant $u_i \in U$ to all task-task sets $TT$
3: compute the optimal path of each TSP in participant-task set $PT$, ie., the path ← *Christofides*(participant-task set)
4: construct the flow network $G = (V, E, C, W)$
5: initialize flow $f$ to 0
6: **while** there exists an augmenting path in the residual network $G_f$ **do**
7: select the augmenting path $p^*$ with minimum cost
8: $c_f(p^*) = q$
9: augment flow $f$ along $p^*$ with $c_f(p^*)$
10: **return** $f$
11: **output** participant set $P$ and accomplished task set $T'$

The MT-MCMF algorithm includes two parts: the one is constructing the flow network of the MCMF model (as shown in line 1-4); the other is finding the optimal solution in the flow network (as shown in line 5-10). Known that $m$ users in the user set $U$ and $n$ tasks in the task set $T$, MT-MCMF starts with combining tasks from all $n$ tasks in task set $T$. Each combined task set contains $q$ tasks, named as the task-task set $TT$. Clearly, there are a total of $C_n^q$ task-task sets $TT$. Then, a total of $m \times C_n^q$ TSPs are formed by adding every participant to all task-task sets, stored in the participant-task sets $PT$. Notice that the optimal path with minimum distance of each TSP is computed by the *Christofides* method [18]. Based on the previous assumptions and results, we construct the flow network $G = (V, E, C, W)$, where $C$ is the capacity of each edge, and $W$ is the total cost of each edge. Subsequently, MT-MCMF finds the optimal solution in the flow network, and the procedure is as follows. First initializing flow $f$ to 0, and then greedily selecting the augmenting path $p^*$ from $s$ to $t$ with the minimum cost in the residual network $G_f$. Augmenting flow $f$ with $C_f(p^*)$ along $p^*$, until there is no augmenting path in the residual network $G_f$. Note that the residual capacity of $p^*$, $C_f(p^*)$, is equal to $q$ because that the flow of the edge between the participant node and the task set node can be 0 or $q$. Finally, outputting the participant set $P$ and the accomplished task set $T'$.

For FPMT, the time complexity of MT-MCMF is $O(mC_n^q q^3)+O(mq(m+n+C_n^q))$. $O(mC_n^q q^3)$ denotes the time complexity of TSPs to compute the short paths by *Christofides*, and $O(mq(m+n+C_n^q))$ is the time complexity of MCMF model to find the optimal solution in the flow network. Obviously, the time complexity of MT-MCMF will increase dramatically with the increasing of $n$ owing to $C_n^q$ task sets.

### MTP-MCMF Algorithm

Finding the optimal solution for the FPMT optimization problem by exploring all possible combinations of tasks is very challenging and time consuming. We thus propose MTP-MCMF with exploring some of possible combinations to decrease the time complexity. Similar to MT-MCMF, MTP-MCMF also includes two parts: the one is constructing the flow network; the other is finding the optimal solution in the flow network. It is clear that there are a total of $C_n^q$ task set nodes in the flow network of MT-MCMF, and it is the main factor leading to the high time complexity. Therefore, the MTP-MCMF algorithm constructs the new flow network with selecting some of task set nodes. Specifically, the edges between the participant nodes and the task set nodes are pruned, and each participant node could connect some of task set nodes instead of all of task set nodes.

For each participant $u_i$, MTP-MCMF first computes the distances between $u_i$ and all $n$ tasks in the task set $T$. Then it combines different tasks with $k$ tasks nearest to $u_i$ ($k \leqslant n$), and each combined task set contains $q$ tasks, named as the task-task set $TT$. Clearly, there are a total of $C_k^q$ task-task sets $TT$. Subsequently, $u_i$ is added to all $C_k^q$ task-task sets to form the participant-task sets $PT$. Therefore, there are a total of $m \times C_k^q$ TSPs stored in $PT$. Similarly, the optimal path with minimum distance of each TSP is computed by *Christofides* method, and the flow network $G'=(V', E', C', W')$ is constructed based on the previous results. Finally, the MTP-MCMF algorithm finds the optimal solution in the flow network, and the procedure is identical to MT-MCMF.

## TASK ALLOCATION OF MPFT PROBLEM

### Problem Analysis

With rich participant resources regarding to tasks, the aim of MPFT is to find the set of participants to complete all tasks with minimizing total incentive payments and the total movement distance. It differs FPMT from the following aspects. First, all tasks should be accomplished and each participant is asked to perform one task to enhance the quality of sensing. Second, we do not need to obtain the precise location of each participant as to urgent tasks. Rather, the participants define preferred working areas in consideration of privacy. Finally, the platform has opportunity to select the participants with lower incentive to reduce the cost. Without loss of generality, we suppose that the incentive of each participant is inversely proportional to the number of users in the area. It is clear that minimizing the total incentive payment is the optimization goal for task publishers to reduce the cost, and minimizing total movement distance is also the optimization objective of MPFT to complete tasks as soon as possible.

Given a set of pre-specified working areas, denoted by $A=\{A_1,A_2...A_i...A_m\}$, and each area includes a set of registered participants according to their interests, $A_i=\{u_1,u_2,u_3...\}$. Moreover, there are a lot of sensing tasks, denoted by $T=\{t_1,t_2...t_j...t_n\}$, and each task $t_j$ will be assigned to $p_j$ users. Notice that $D_{ij}$ is the distance between the participant in area $A_i$ and task $t_j$, $C_i$ is the incentive of the participant in area $A_i$. In addition, $x_{ij}$ is the number of participants in area $A_i$ accomplishing the task $t_j$. Then, the MPFT problem can be formulated as Eq. (5)-(9):

$$\text{minimize } \sum_{i=1}^{m} C_i \times \sum_{j=1}^{n} x_{ij} \quad (5)$$

$$\text{minimize } \sum_{i=1}^{m} \sum_{j=1}^{n} D_{ij} \times x_{ij} \quad (6)$$

Subject to:

$$\sum_{j=1}^{n} x_{ij} \leq |A_i| \ (1 \leq i \leq m) \quad (7)$$

$$\sum_{i=1}^{m} x_{ij} = p_j \ (1 \leq j \leq n) \quad (8)$$

$$x_{ij} \in Z^n \quad (9)$$

The main challenge of MPFT relates to the two contradictory optimization objectives. Therefore, we use the multi-objective optimization model to solve the MPFT problem. In general, the main idea to solve the multi-objective optimization problem is transforming multi-objective to single-objective, so we use the linear weight method [21] and constrain method [19] to achieve it.

### The Linear Weight Method

The main idea of the linear weight method is adding up all objective functions of the multi-objective optimization problem to the comprehensive objective function by assigning different weights. Note that values of incentive and distance are two different dimensions, and we should not add up the values simply. There are two steps to transform MPFT to the single-objective optimization problem by the linear weight method: the first one is scaling the values of original objective functions; the second one is determining the weights. We adopt the following scale model in the [1], as shown in Eq. (10). $f(x)$ is the value of objective function, $f^{max}$ and $f^{min}$ are maximum and minimum values of objective function $f(x)$ respectively.

$$f'(x) = \frac{f(x)-f^{min}}{f^{max}-f^{min}} \quad (10)$$

We use two variables, $k_1$ and $k_2$, to represent the weights of incentive and distance respectively owing to some unpredictable requirements. Therefore, the objective function of MPFT is transformed as Eq. (11):

$$\text{minimize } k_1 \times \frac{\sum_{i=1}^{m} C_i \times \sum_{j=1}^{n} x_{ij} - C_{min}}{C_{max} - C_{min}} + k_2 \times \frac{\sum_{i=1}^{m} \sum_{j=1}^{n} D_{ij} \times x_{ij} - D_{min}}{D_{max} - D_{min}} \quad (11)$$

Obviously, the MPFT problem is the integer linear programming (ILP) [7], and we use the branch-and-bound method [20] of ILP to obtain the optimal solution. Based on the above analyses, we propose W-ILP algorithm to solve the MPFT problem.

W-ILP contains two parts: the one is computing maximum and minimum values of incentive and distance, and generalizing the new single-objective optimization problem by the linear weight method; the other is solving the problem by the branch-and-bound method. W-ILP picks the participants with lower incentive greedily to minimize the total incentive. Correspondingly, the total distance tends to reach a maximum value. In addition, each task should be performed by the nearest participants to minimize the total distance without considering the incentive. Based on the above results, the single-objective optimization problem is formed. Then we view the single-objective optimization problem as the integer linear programming problem, and obtain the optimal solution by branch-and-bound method. W-ILP first forms a rooted tree with the original optimal solution of relaxation problem of ILP problem, and ascertains the upper and lower bounds according to current solution. Then, it greedily adjusts the original optimal solution until satisfying the integer constraints of variables, which includes branch, bound and pruning three steps. Finally, W-ILP outputs the number of participants in each area for each task.

### The Constraint Method

It is a good way to solve MPFT by transforming to the single-objective optimization problem. For the linear weight method, its implement is simpler and the weights of different objective functions are controllable according to the different constraints. However, there are still some disadvantages, such as the optimal degree of solutions cannot be guaranteed. Additionally, we transform MPFT to the single-objective optimization problem by the constraint method. In short, the constraint method chooses one of objective functions as the optimization goal, and the residual objective functions are expressed as the constraint conditions by adding some bounds. There are two objective functions in MPFT, one is the total incentive, and the other is the movement distance. For task publishers, the total incentive budget is controllable. Therefore, we set the incentive budget $C$ to the total incentive, and MPFT is transformed as Eq. (12)-(13):

$$\text{minimize } \sum_{i=1}^{m} \sum_{j=1}^{n} D_{ij} \times x_{ij} \quad (12)$$

$$\sum_{i=1}^{m} C_i \times \sum_{j=1}^{n} x_{ij} \leq C \quad (13)$$

We propose C-ILP algorithm to solve the MPFT problem based on the constraint method. Similarly, we use branch and bound method to obtain the optimal solution of ILP.

## EVALUATION

### Dataset and Experiment Setups

*Dataset*

The dataset we used in evaluation is the D4D dataset [2], which contains two data types. One data type contains the information about cell towers, including tower id, latitude and longitude. The other one contains 50,000 users' phone call records in Ivory Coast. Thus, we design experiments based on the location of cell towers and phone users. We set the task information about location based on the cell towers. Besides, the location of participant is cell tower's location where this participant made the phone call currently.

*Experiment setups*

For FPMT, the number of participants needed to finish one task is at most $p$. Considering that the value of $p$ makes no difference to experimental results, we set it to 6 in the following experiments. In addition, in view of the completion of tasks, the number of tasks that each participant can perform at one time is set randomly between 2 and 7. To measure the completion time of one participant knowing the movement distance, we simply assume that the average moving speed of one participant by walking is 70 meters per minute. In each experiment, the locations of the tasks are randomly distributed to a group of cell towers within a target area, and participants to be selected are users in D4D who make phone calls around these cell towers within a pre-specified time window (tasks are also assumed published during that period). Moreover, to measure whether our proposed methods perform well under different spatial distributions of tasks, we adopt three types of task distributions.

- ***Compact distribution*** - the tasks are relatively close to each other and all of them are distributed mainly within a portion of the user area.
- ***Scattered distribution*** - the tasks are scattered over the target area which is wider than the user area.
- ***Hybrid distribution*** - the tasks are distributed in the user area randomly.

An illustration of the three types of distributions used in our experiments is given in Fig. 3.

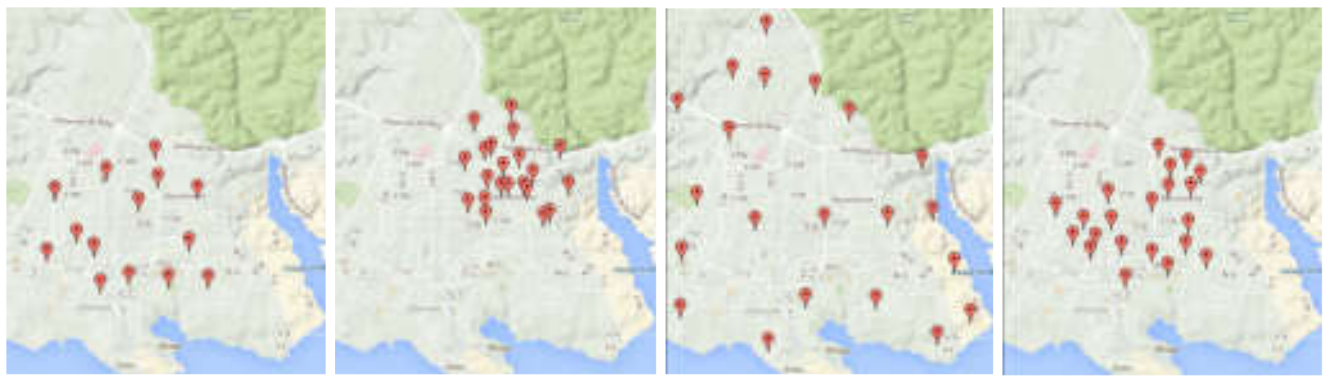

(a) users (b) compact (c) scattered (d) hybrid

**Figure 3. An illustration of the three types of task distribution.**

For MPFT, the precise location of every participant is unknown because of privacy, and it is replaced by the location of areas. In our experiments, the entire area of users is divided into 6 parts distributed evenly among the city. The number of participants in each area is randomly generated between 10 and 100. We assume that the incentive of each participant in 6 different kinds of areas is between 1 and 10, and it is inversely proportional to the number of participants in the areas. In addition, in view of MPFT with few tasks, we assume that there are 20 tasks randomly distributed in the city, and the number of participants needed by each task is set to 5 to ensure the quality of sensing, as shown in Fig. 4.

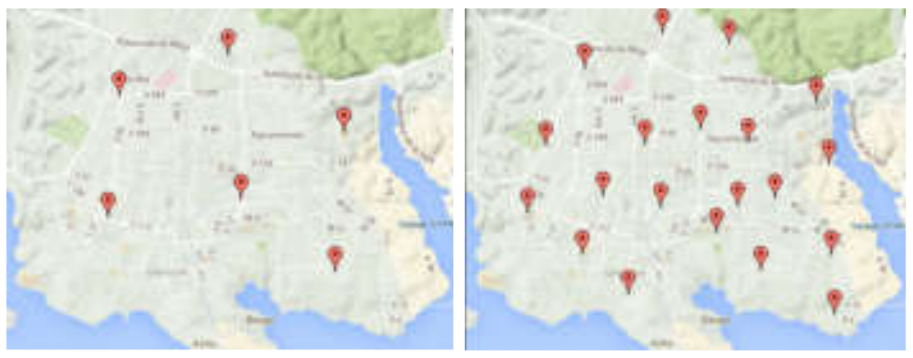

(a) areas (b) tasks

**Figure 4. Distributions of areas and tasks.**

### Baseline Methods and Evaluation Metrics

#### *Baseline methods*

In our evaluations, we provide one baseline method for FPMT named MT-GrdPT, and two baseline methods for MPFT named W-Grd and C-Grd.

- **MT-GrdPT** - MT-GrdPT algorithm aims to minimize the total movement distance based on maximizing the number of accomplished tasks. MT-GrdPT greedily picks the tasks for each participant with minimum distance between the participants and tasks, until there are $q$ tasks in the task set for each participant. Note that the participant moves from previous location to the location of allocated task.
- **W-Grd** - The basic idea of the W-Grd algorithm is to simply select the participants and tasks with minimum value of the objective function. Notice that W-Grd transforms the bi-objective problem to the single objective problem based on the linear weight method (the detail is introduced in section 3).
- **C-Grd** - C-Grd algorithm tries to select the participant set and corresponding task set with minimum distance while satisfying the incentive budget based on the constraint method. C-Grd first picks the participants and tasks having minimum distance as preliminary selected set in each iteration. Then, C-Grd computes whether the incentive of the preliminary selected set satisfies the incentive budget. If it meets, the preliminary selected set is the final participant and task set. Otherwise, C-Grd needs to adjust the assignment of selected set between participants and tasks constantly, until its total incentive satisfies the incentive budget.

#### *Evaluation Metrics*

For FPWT, the number of accomplished tasks and the total movement distance of selected participants are the chief indicator for comparison. In addition, for some urgent tasks in FPWT, we also need that the average completion time of $q$ tasks is as short as possible. Finally, the running time of algorithm is also important for solving the problem.

For MPFT, the total incentive payments of participants are significant to task publishers in the MCS platform. Moreover, the total movement distance of accomplishing tasks is another important indicator.

### Evaluation of FPMT

The optimization goal of FPMT is to minimize the total distance based on maximizing the number of accomplished tasks. Note that each of $m$ participants is required to perform $q$ tasks. Therefore, the maximum number of accomplished tasks is $m \times q$. In our following experiments, the number of accomplished tasks is no longer an important factor for comparison because that three numbers of accomplished tasks obtained by three methods reach maximal values. We compare the performance of three algorithms (like the total distance, the completion time) under different situations, such as the different number of tasks, the various distributions of tasks, and so on.

#### *Different number of tasks*

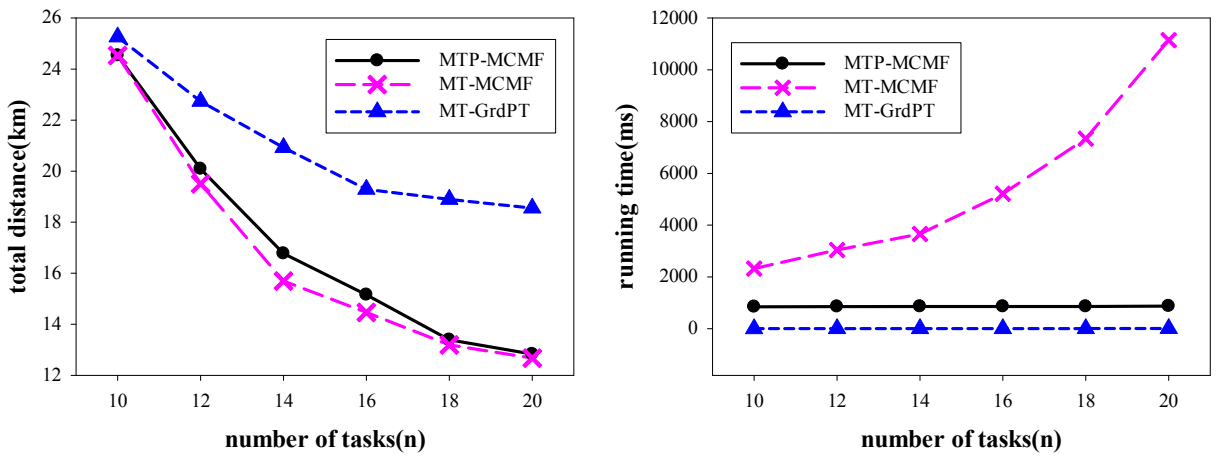

(a) (b)

**Figure 5. Performance comparison under different number of tasks.**

In Fig. 5, we present the performance comparison on the total distance and running time among three methods under different number of tasks. In addition, we assume that there are 10 users in the MCS platform volunteering to perform sensing tasks, and each one is required to accomplish 5 tasks. From Fig. 5(a), we can see that the total movement distance decreases monotonically with the increasing number of tasks, because there may be a better choice among the newly added tasks for participants. It is clear that MTP-MCMF and MT-MCMF outperform MT-GrdPT in all different number of tasks, and the performance of MT-GrdPT is getting worse with the increasing number of tasks, as shown in Fig. 5(a). Moreover, we can find that the result of MTP-MCMF is close to MT-MCMF. MTP-MCMF picks 10 tasks from $n$ tasks to form the task sets leading to lose the optimal solution possibly, and the running time of MTP-MCMF remain virtually unchanged. However, MT-MCMF forms the task sets using all of $n$ tasks, and it causes dramatic increase of running time with increasing number of tasks, as shown in Fig. 5(b).

#### *Different value of k*

Here we compare the performance of MTP-MCMF with different value of $k$ (combining different tasks with $k$ tasks nearest to the user). Note that $k$ is no more than the number of tasks ($k \leqslant 15$). There are 15 sensing tasks and 10 participants in the MCS platform, and one participant needs to perform 5 tasks. Figure 6(a) illustrates that MTP-MCMF makes the total distance reduce gradually with the increase of $k$. For MT-MCMF, it chooses the optimal solution from all $C_{15}^{5}$ possible task sets causing the longer running time, as shown in Fig. 6 (b). Nevertheless, MTP-MCMF selects the better solution in $C_{k}^{5}$ task sets which is a part of all task sets, and this may lead to lose the optimal solution. It is clear that the distance between MTP-MCMF and MT-MCMF is equal when the value of $k$ is 12. However, the running time of MT-MCMF is about three times as much as MTP-MCMF. The results indicate that there is no need to take into account all possible task sets, which could waste storage space and runtime. Moreover, we can conclude that MTP-MCMF can approximate to the optimal solution of MT-MCMF with the appropriate value of $k$.

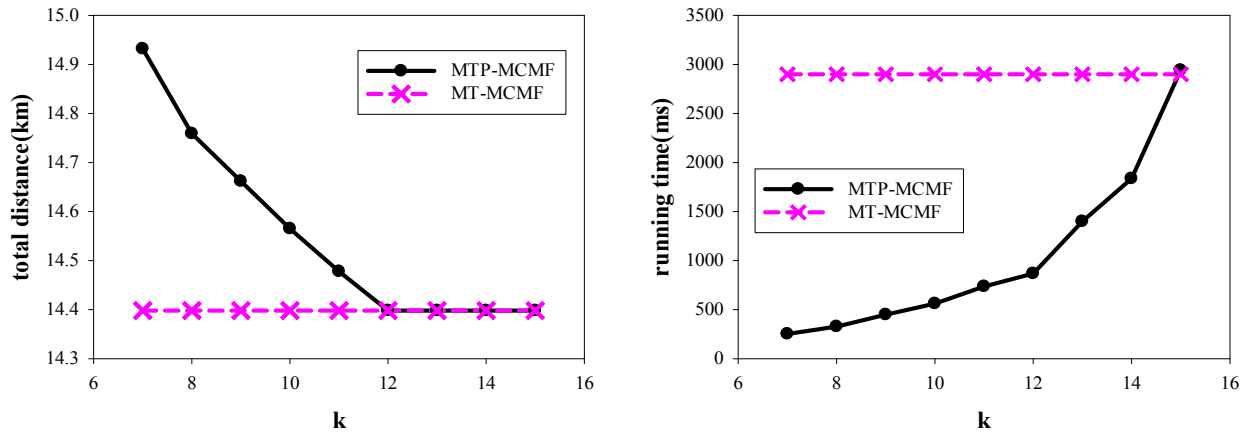

(a) (b)

**Figure 6. Performance of MTP-MCMF.**

It is clear that the performance of the MTP-MCMF highly depends on the value of $k$. We thus present the appropriate value of $k$ of the MTP-MCMF under different number of tasks and different value of $q$. From Fig. 7(a), we can see that the value of $k$ increases first and then keeps stable with the increasing of the number of tasks. The reason is that participants tend to perform tasks nearby when there are sufficient number of tasks. For Fig. 7(b), the participant is selected to complete $q$ tasks from $k$ tasks by MTP-MCMF. Therefore, the value of $k$ increases constantly with the increasing of $q$, and it is about twice of the value of $q$.

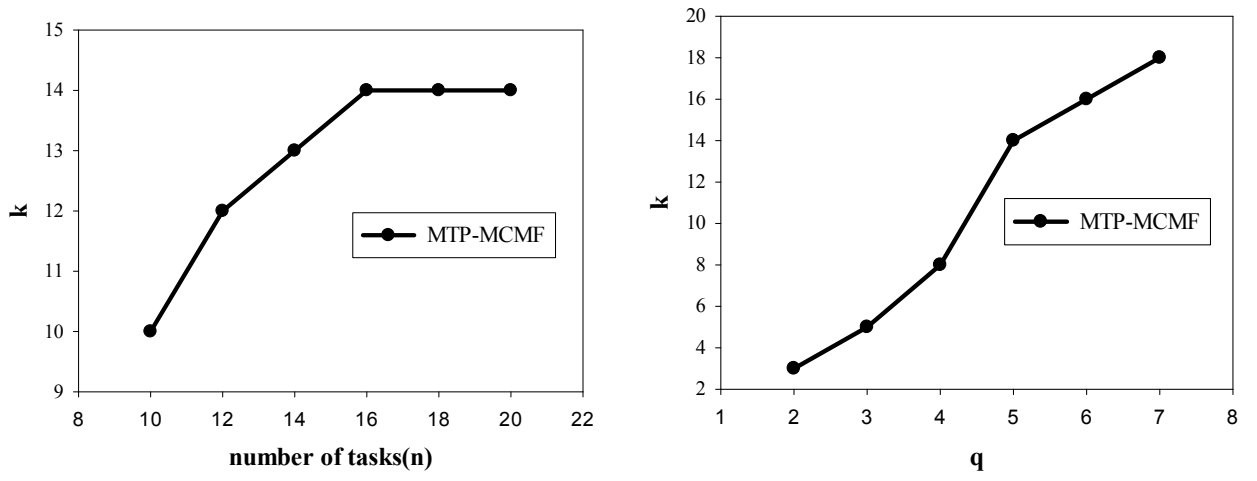

(a) (b)

**Figure 7. Appropriate value of *k* of the MTP-MCMF.**

*Different value of q*

We assume that each participant is required to perform $q$ tasks, now we compare the task allocation by three algorithms under the different number of tasks accomplished by one participant. Figure 8(a) presents that the total distance increases constantly owing to the increasing number of accomplished tasks. We can see that MT-GrdPT performs worse than other two algorithms in the total distance, and the difference becomes obvious with the increase of $q$. MT-GrdPT is the greedy algorithm receiving the locally optimal solution easily in a relatively short time that leads to the final result get worse with the increasing number of iterations. MTP-MCMF and MT-MCMF have the similar result of the total distance, because we choose the appropriate value of $k$ according to previous experimental results. Therefore, MT-MCMF costs more running time than MTP-MCMF, as shown in Fig. 8 (b). In general, for some urgent tasks, it is a good way to reduce the completion time by adjusting the number of completed tasks of each participant.

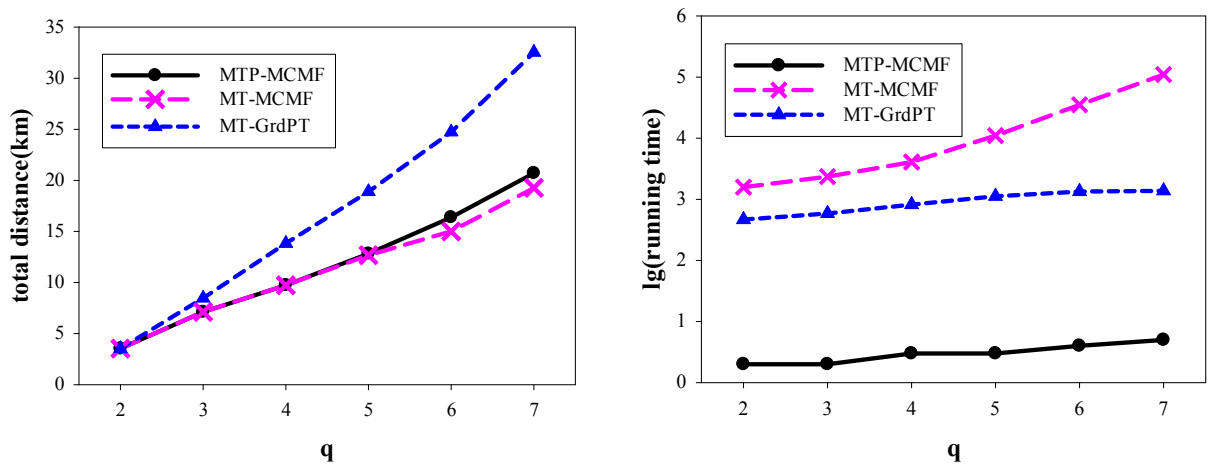

(a) (b)

**Figure 8. Performance comparison under different value of *q*.**

*Different number of participants*

We study how the proposed algorithms perform when there are different number of participants in the MCS platform. We assume that there are 20 sensing tasks and each participant in the platform is required to perform 5 tasks. As shown in Fig. 9(a), it is clear that the total distance grows steadily with the increasing number of participants. The total number of accomplished tasks is in direct proportion to the number of participants, and the more the accomplished tasks, the longer distance participants move. In Fig. 9(b), we can see that the completion time of 5 tasks remains about the same because of unchanged number of tasks, and a participant needs to spend about 20-30 minutes to perform 5 tasks under this kind of task distribution.

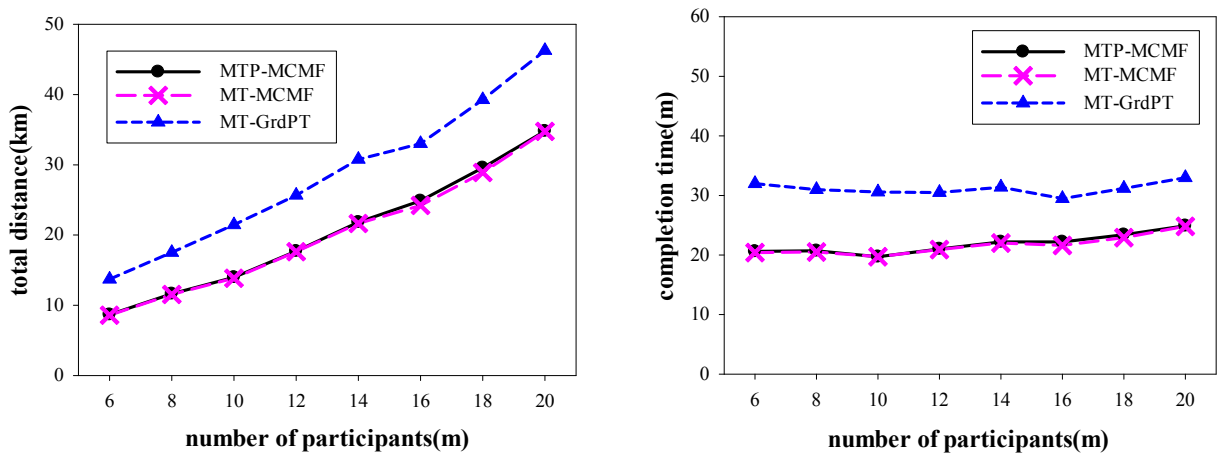

(a) (b)

**Figure 9. Performance comparison under different number of participants.**

*Different task distributions*

Now we investigate the question which is beneficial to select participants accomplishing tasks in three situations: compact distribution, hybrid distribution, and scattered distribution. In Fig. 10(a), we can find that the scattered distribution makes the total distance longest of all three task distributions, and the compact distribution has the shortest distance. Because the distance between tasks under the scattered distribution is longer than other two task distributions, and participants need to move the longer distance and spend more time to accomplish all tasks, as show in Fig. 10(b) and Fig. 11(c). This means that the total distance and the completion time are affected by different task distributions. Specifically, MTP-MCMF and MT-MCMF outperform MT-GrdPT of all task distributions.

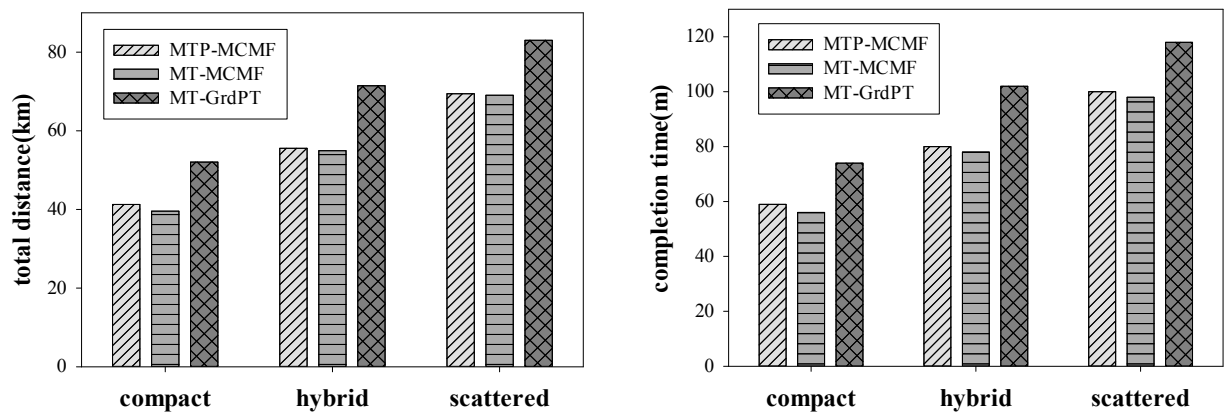

(a) (b)

**Figure 10. Performance comparison under different task distributions.**

We also study the completion of each task by MTP-MCMF under three task distributions, as shown in Fig. 11. We assume that each task to be performed by at most 6 participants, and each participant needs to accomplish 5 tasks. We suppose that there are 12 sensing tasks and 8 participants separately under three task distributions. As shown in Fig. 11(a), the number of participants for each task is heterogeneous under the compact and scattered distribution relative to the hybrid distribution. In Fig. 11(b), we show the overall situation of task completion by calculating the variance of task set based on the number of participants under four situations. '8P12T' means that the number of participants is 8 and the number of tasks is 12. We can find that the variance of the hybrid distribution is smallest of three task distributions under all four situations, and it means the number of participants for each task is no big difference among all the tasks. The hybrid distribution of tasks is similar with the distribution of user's areas. In general, the difference between the distribution of tasks and participants is key factor influencing the overall situation of task completion.

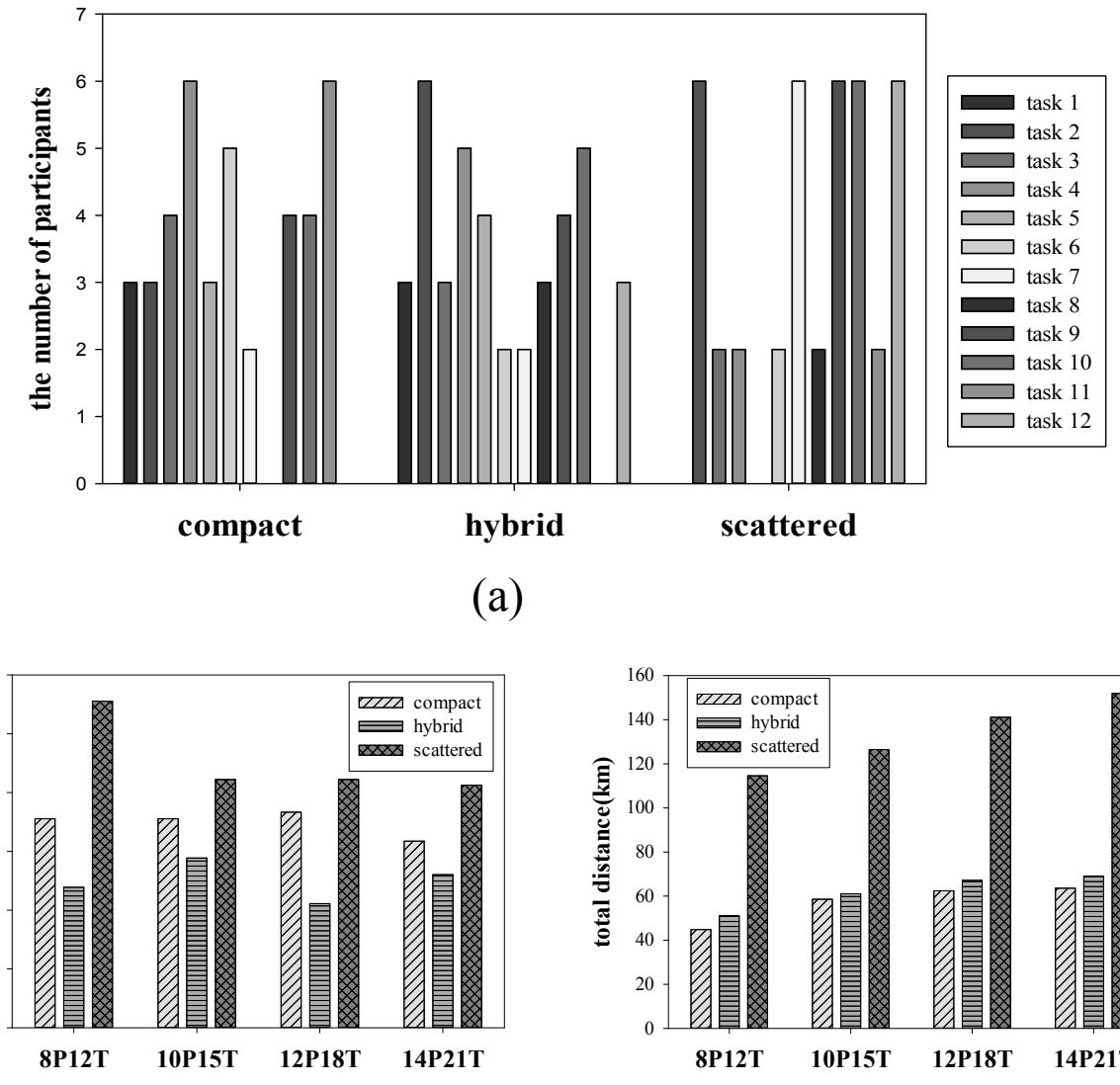

(b) (c)

**Figure 11. Completion of tasks.**

### Evaluation of MPFT

The optimization goals of MPFT are to minimize the total incentive and the total movement distance. For the bi-objective optimization problem, there are some groups of non-inferior solutions instead of the global optimal solution, known as the Pareto solution [26]. Therefore, we study the results of proposed algorithms in terms of both the total incentive and the total movement distance.

#### *The Linear Weight Method*

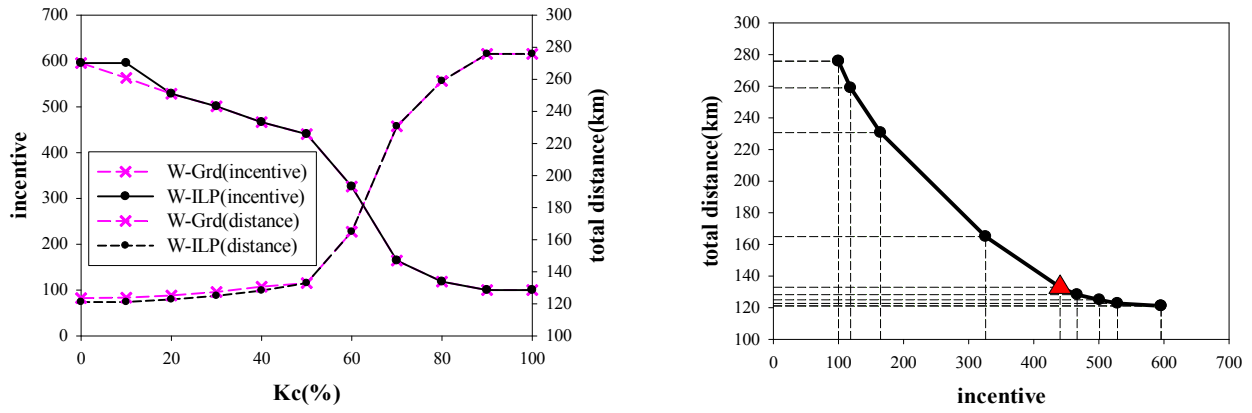

(a) (b)

**Figure 12. Performance comparison under different weights.**

Figure 12 shows the incentive and total distance of W-Grd (greedy algorithm based on the linear weight method) and W-ILP (integer linear programming based on the linear weight method) algorithms under different weighted values. $K_c$ is the weight of incentive, and $K_d$ is the weight of the total distance. Clearly, the sum of $K_c$ and $K_d$ is 1. As shown in Fig. 12(a), the incentive decreases gradually to the minimum value with the increase of $Kc$, but the total distance increases to the maximum value. Specially, the result of W-Grd is close to W-ILP, which implies that the performance of W-Grd is close to the optimal solution. From Fig. 12(b), we can see nine groups of the Pareto solutions of W-ILP. It is clear that it is impossible to find a solution with the minimum distance and minimum incentive. From the perspective of the incentive, the first group of non-inferior solutions is a relatively good choice. On the contrary, the last group of non-inferior solutions is a relatively good choice in terms of the distance. In general, the fifth group of non-inferior solutions (triangular symbols in Fig. 12(b)) is not a bad solution for both the incentive and distance. These results suggest the tradeoff between the incentive cost and distance when deciding the weighted values of $K_c$ and $K_d$.

#### *The Constraint Method*

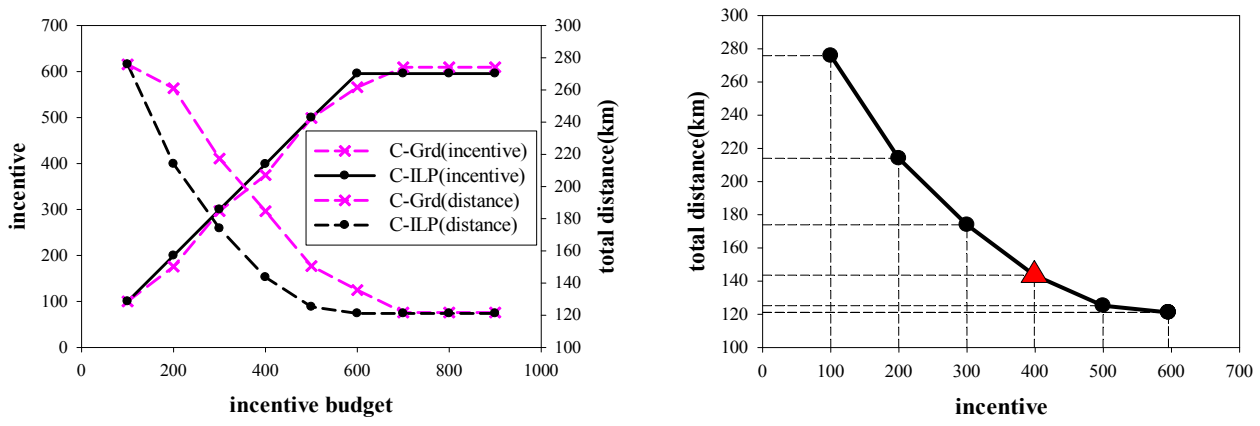

(a) (b)

**Figure 13. Performance comparison under different incentive budgets.**

In Fig. 13, we present the performance comparison on the incentive and distance between C-Grd (greedy algorithm based on the constraint method) and C-ILP (integer linear programming based on the constraint method). It is clear that C-ILP outperforms C-Grd method in all incentive budgets. We can find that the total distance of C-Grd is close to the C-ILP, but C-Grd has higher incentive than C-ILP, as shown in Fig. 13(a). Figure 13(b) shows that there are six groups of non-inferior solutions of the C-ILP algorithm with the different incentive budgets. Like the previous result, the lower incentive, the longer total distance. However, the difference is that the distribution of non-inferior solutions is uniform and controllable in the constraint method. For example, the MCS platform could select a group of solutions with the lower incentive under limited incentive budget.

The advantage of the linear weight method is that the weights of two objective functions (i.e., the movement distance and the incentive cost) are adjustable by the task publisher under different situations. For example, the weight of the movement distance should be higher than the incentive to reduce the movement time in consideration of the emergency of tasks. In general, it is difficult to set the appropriate value of weight to guarantee the optimal degree of solutions, the reason is that a little difference between two values of weights could result in significant performance difference. The constraint method can easily get the optimal results given the incentive budget.

## DISCUSSIONS

This subsection discusses issues that are not reported or addressed in this work due to space and time constraint, which can be added to our future work.

***Diversity of participants.*** Each participant is required to complete the same number of sensing tasks in view of the fairness in FPMT, which do not take into account the diversity of different participants. For example, some users have ability to perform more tasks to improve the completion of sensing tasks. In MPFT, we obtain the number of participants in each area, and the participant makes their own decisions to pick one task to complete, which could leads to some contradictions. Hence, we will select the unique participant for each task considering the character of each participant in our future work.

***Various constraints.*** We take into account the two general constraints when defining FPMT and MPFT problems. However, in many case, there are other types of constraints to be considered, such as the interests of participants, the budget of MCS platforms, etc. Therefore, in our future work, we plan to improve the practicality of our proposed framework by considering other constraints.

***Incentive models.*** For MPFT, we consider that the incentive of each participant is inversely proportional to the number of users in the area, which do not take into account other factor (e.g., the movement distance). In addition, the objective of MPFT is to minimize the total incentive cost, which could diminish the participant's enthusiasm for long-term runnings. As for future work, we will define the incentive payment models based on different situations, and make a balance between the incentive cost of the MCS platform and the participants' long-term interest.

***Methods for task allocation.*** In our paper, we use the MCMF model to solve FPMT. However, one disadvantage is that we need to calculate all possible task combinations for each participant, which limits the scale of tasks and increases the computation cost. Our proposed algorithms are flexible by considering other constraints (e.g., budget of MCS platforms), but they may not be applicable in view of some complicated constraints (like the quality of completed tasks). We intend to improve our algorithms to adapt to different situations.

***Large-scale user study.*** In this work, the evaluation of our algorithms is based on a real-world human mobility dataset, and we distribute tasks according to different schemes with certain assumptions. We are now collaborating with the local government to build a city-level MCS platform for citizens to report municipal problems (e.g., road collapse, public facility damage, and noise disturbance), urban dynamics (e.g., traffic jams, accidents), etc. We intend to introduce our multi-task allocation framework in the platform and make large-scale user studies over it, which will help identify practical issues and improve our framework.

## CONCLUSION

In this paper, we proposed two bi-objective optimization problems for participant selection, the one is the FPMT problem with the deficient participant resources, and the other is the MPFT problem with the rich participant resources. We adopted the MCMF model to solve the FPMT problem, and proposed the MT-MCMF and MTP-MCMF algorithms to select participants with the maximum number of accomplished tasks and minimum total traveling distance. In addition, we used the multi-objective optimization model to solve the MPFT problem, and proposed the W-ILP and C-ILP algorithms to select participants with the minimum total incentive payments and minimum total movement distance. Evaluations based on a large-scale real-world dataset showed that our proposed algorithms of two problems outperform the baseline approaches. As for future work, we will consider other factors that may affect the selection of participants in multi-task MCS environments. New optimization methods and theoretical foundations will also be studied and leveraged.

## ACKNOWLEDGEMENT

This work was partially supported by the National Basic Research Program of China (No.2015CB352400), the National Natural Science Foundation of China (No. 61332005, 61373119).

## REFERENCES

1. M. Alrifai, D. Skoutas, and T. Risse. 2010. Selecting skyline services for QoS-based web service composition. In *proceedings of the 19th international conference on World Wide Web*, 11-20.
2. V. D. Blondel, M. Esch, C. Chan, F. Clerot, P. Deville, E. Huens, F. Morlot, Z. Smoreda, and C. Ziemlicki. 2012. Data for development: the d4d challenge on mobile phone data. arXiv preprint arXiv:1210.0137.
3. M. Buhrmester, T. Kwang, and S. D. Gosling. 2011. Amazon's mechanical turk a new source of inexpensive, yet high-quality, data? *Perspectives on Psychological Science*, 6: 3-5.
4. Campaignr: http://research.cens.ucla.edu/urban/
5. G. Cardone, L. Foschini, P. Bellavista, A. Corradi, C. Borcea, M. Talasila, and R. Curtmola. 2013. Fostering participaction in smart cities: a geo-social crowdsensing platform. *IEEE Communications Magazine*, 51: 112-119.
6. M. H. Cheung, R. Southwell, F. Hou, and J. Huang. 2015. Distributed Time-Sensitive Task Selection in Mobile Crowdsensing. In *Proceedings of the 16th ACM International Symposium on Mobile Ad Hoc Networking and Computing* (MobiHoc '15), 157-166.
7. J. Choi, H. Seo, and C. Lee. 2013. An integer linear programming (ILP)-based optimization for finding the optimal independent sets in wireless ad hoc networks. In *Proceedings of the 7th International Conference on Ubiquitous Information Management and Communication* (ICUIMC '13).
8. C. Click and K. D. Cooper. 1995. Combining analyses, combining optimizations. *ACM Transactions on Programming Languages and Systems*, 17: 181-196.
9. V. Coric and M. Gruteser. 2013. Crowdsensing Maps of On-street Parking Spaces. In *Proceedings of the 2013 IEEE International Conference on Distributed Computing in Sensor Systems* (DCOSS '13), 115-122.
10. T. Das, P. Mohan, V. N. Padmanabhan, R. Ramjee, and A. Sharma. 2010. PRISM: platform for remote sensing using smartphones. In *Proceedings of the 8th international conference on Mobile systems, applications, and services* (MobiSys '10), 63-76.
11. R. K. Ganti, F. Ye, and H. Lei. 2011. Mobile crowdsensing: Current state and future challenges. *IEEE Communications Magazine*, 49: 32-39.
12. B. Guo, H. Chen, Z. Yu, X. Xie, S. Huangfu, and D. Zhang. 2015. FlierMeet: A Mobile Crowdsensing System for Cross-Space Public Information Reposting, Tagging, and Sharing. *IEEE Transactions on Mobile Computing*, 14: 2020-2033.
13. B. Guo, Z. Wang, Z. Yu, Y. Wang, N. Y. Yen, R. Huang, and X. Zhou. 2015. Mobile Crowd Sensing and Computing: The Review of an Emerging Human-Powered Sensing Paradigm. *ACM Computing Surveys*, 48: 7.
14. M. Hadji and D. Zeghlache. 2012. Minimum Cost Maximum Flow Algorithm for Dynamic Resource Allocation in Clouds. In *Proceedings of the 2012 IEEE Fifth International Conference on Cloud Computing* (CLOUD '12), 876-882.
15. S. He, D. Shin, J. Zhang, and J. Chen. 2014. Toward optimal allocation of location dependent tasks in crowdsensing. In *Proceedings of 2015 IEEE Conference on Computer Communications* (INFOCOM '14), 745-753.
16. J. Hershberger, S. Suri, and A. Bhosle. 2007. On the difficulty of some shortest path problems. *ACM Transactions on Algorithms*, 3: 5.
17. J. Hicks, N. Ramanathan, D. Kim, M. Monibi, J. Selsky, M. Hansen, and D. Estrin. 2010. AndWellness: an open mobile system for activity and experience sampling. In *Proceedings of Wireless Health* (WH '10), 34-43.
18. S. Nallaperuma, M. Wagner, F. Neumann, B. Bischl, O. Mersmann, and H. Trautmann. 2013. A feature-based comparison of local search and the christofides algorithm for the travelling salesperson problem. In *Proceedings of the twelfth workshop on Foundations of genetic algorithms XII* (FOGA XII '13), 147-160.
19. D. S. Prasanna, G. Garima, K. Ety, and S. Manivannan. 2011. Multi-objective optimization of cylindrical fin heat sink using Taguchi based ε-constraint method. In *proceedings of Internation Conference on Sustainable Energy and Intelligent Systems* (SEISCON 2011), 678 -682.
20. J. Qi, Z. Xu, Y. Xue, and Z. Wen. 2012. A branch and bound method for min-dist location selection queries. In *Proceedings of the Twenty-Third Australasian Database Conference* (ADC '12), 51-60.
21. L. Qi, Y. Tang, W. Dou, and J. Chen. 2010. Combining Local Optimization and Enumeration for QoS-aware Web Service Composition. In *Proceedings of the 2010 IEEE Conference on Web Services* (ICWS '10), 34-41.
22. M. Ra, B. Liu, T. F. L. Porta, and R. Govindan. 2012. Medusa: A programming framework for crowd-sensing applications. In *Proceedings of the 10th international conference on Mobile systems, applications, and services* (MobiSys '12), 337-350.
23. S. Reddy, D. Estrin, and M. Srivastava. 2010. Recruitment framework for participatory sensing data collections. In *Proceedings of International Conference on Pervasive Computing*, 138-155.
24. Z. Song, C. H. Liu, J. Wu, J. Ma, and W. Wang. 2014. QoI-Aware Multitask-Oriented Dynamic Participant Selection with Budget Constraints. *IEEE Transactions on Vehicular Technology*, 63: 4618-4632.

25. H. Sugihara, J. Komoto, and K. Tsuji. 2002. A multi-objective optimization model for urban energy systems in a specific area. In *Proceedings of 2014 IEEE International Conference on Systems, Man and Cybernetics*, 5: 6-11.
26. J. Wang, Y. Zhou, H. Yang, and H. Zhang. 2012. A Trade-off Pareto Solution Algorithm for Multi-objective Optimization. In *Proceedings of the Fifth International Joint Conference on Computational Sciences and Optimization* (CSO '12), 123-126.
27. L. Wang, D. Zhang, A. Pathak, C. Chen, H. Xiong, D. Yang, and Y. Wang. 2015. CCS-TA: quality-guaranteed online task allocation in compressive crowdsensing. In *Proceedings of the 2015 ACM International Joint Conference on Pervasive and Ubiquitous Computing* (UbiComp '15), 683-694.
28. K. D. Wayne. 2002. A polynomial combinatorial algorithm for generalized minimum cost flow. *Mathematics of Operations Research*, 27: 445-459.
29. M. Xiao, J. Wu, L. Huang, Y. Wang, and C. Liu. 2015. Multi-Task Assignment for CrowdSensing in Mobile Social Networks. In *Proceedings of the 2015 IEEE Conference on Computer Communications* (INFOCOM '15), 2227-2235.
30. H. Xiong, D. Zhang, G. Chen, L. Wang, and V. Gauthier. 2015. CrowdTasker: Maximizing coverage quality in Piggyback Crowdsensing under budget constraint. In *Proceedings of the 2015 IEEE Conference on Pervasive Computing and Communications* (PerCom '15), 55-62.
31. Z. Yan, N. Banerjee, D. Chakraborty, A. Misra, M. Srivastava, and S. Mittal. 2013. PUCAA: 1st international workshop on pervasive urban crowdsensing architecture and applications. In *Proceedings of the 2013 ACM conference on Pervasive and ubiquitous computing adjunct publication* (UbiComp '13 Adjunct), 1055-1062.
32. D. Zhang, H. Xiong, L. Wang, and G. Chen. 2014. CrowdRecruiter: selecting participants for piggyback crowdsensing under probabilistic coverage constraint. In *Proceedings of the 2014 ACM International Joint Conference on Pervasive and Ubiquitous Computing* (UbiComp '14), 703-714.
33. Y. Zheng, F. Liu, and H. Hsieh. 2013. U-Air: When urban air quality inference meets big data. In *Proceedings of the 19th ACM SIGKDD international conference on Knowledge discovery and data mining* (KDD '13), 1436-1444.